\documentclass[aps,preprint,showpacs,superscriptaddress,groupedaddress]{revtex4}
\pdfoutput=1
\usepackage[pdftex]{color,graphicx}
\usepackage{amsmath,epsfig}
\usepackage{mathrsfs}
\usepackage{amssymb}
\usepackage{mathtools}
\usepackage{graphics}
\usepackage{cancel}
\usepackage{bbm}
\usepackage{slashed}

\newcommand{\be}{\begin{equation}}
\newcommand{\ee}{\end{equation}}
\newcommand{\bea}{\begin{eqnarray}}
\newcommand{\eea}{\end{eqnarray}}

\begin{document}

\title{$N_{\rm eff}$ in low-scale seesaw models versus the lightest neutrino mass}

\author{P. ~Hern\'andez}
\email[]{m.pilar.hernandez@uv.es}
\affiliation{IFIC (CSIC) and Dpto. F\'{\i}sica Te\'orica, \\
Universidad de Valencia, Edificio Institutos Investigaci\'on, \\
Apt.\ 22085, E-46071 Valencia, Spain.}
\author{M.~Kekic}
\email[]{Marija.Kekic@ific.uv.es}
\affiliation{IFIC (CSIC) and Dpto. F\'{\i}sica Te\'orica, \\
Universidad de Valencia, Edificio Institutos Investigaci\'on, \\
Apt.\ 22085, E-46071 Valencia, Spain.}
\author{J. Lopez-Pavon}
\email[]{jlpavon@sissa.it}
\affiliation{SISSA, via Bonomea 265, 34136 Trieste, Italy.}
\affiliation{INFN, sezione di Trieste, 34136 Trieste, Italy.}

\date{\today}

\begin{abstract}
We evaluate the contribution to $N_{\rm eff}$ of the extra sterile states in low-scale Type I 
seesaw models (with three extra sterile states). We explore the full parameter space and find
 that at least two of the heavy states always reach thermalization in the Early 
Universe, while the third one might not thermalize provided the lightest neutrino mass is below ${\mathcal O}(10^{-3}$eV). 
Constraints from cosmology therefore severely restrict the spectra of heavy states  in the range
1 eV- 100 MeV. The implications for neutrinoless double beta decay 
are also discussed. 
\end{abstract}

\preprint{IFIC/14-25\\ SISSA 19/2014/FISI}

\pacs{14.60.St, 98.80.Cq}
\maketitle

\section{Introduction}

The simplest extension of the Standard Model (SM) that can account for the observed neutrino masses is a Type I seesaw model \cite{seesaw} with $N\geq 2$ extra singlet Majorana fermions. 
The Majorana masses, that we globally denote as $M$, constitute a new scale of physics (the seesaw scale) which is presently unknown. Since the light neutrino masses are a combination of the Yukawa couplings, the electroweak scale and the seesaw scale, the latter can be arbitrary if the Yukawas are adjusted accordingly. 
As a result, the seesaw scale is presently unconstrained to lie anywhere above ${\mathcal O}(eV)$ up to ${\mathcal O}(10^{15}$ 
GeV)\footnote{The range $~10^{-10}$eV-1eV can be excluded from oscillation data \cite{miniseesaw,Donini:2011jh}, while a scale above $10^{15}$GeV would require non-perturbative Yukawa couplings. }.  The determination of this scale is one of the most important open questions in neutrino physics. 

It is often assumed that the seesaw scale is very high, above the electroweak scale. However, in the absence 
of any other hint of new physics beyond the SM, the possibility that the seesaw scale could be at the 
electroweak scale or lower should be seriously considered. As far as naturalness goes, the model with a low-scale is technically natural, since in the limit $M\rightarrow 0$, a global lepton number symmetry is recovered: neutrinos becoming Dirac particles by the pairing of the Majorana fermions. 

The spectrum of $N=3$ Type I seesaw models contains six Majorana neutrinos: the three lightest neutrinos, mostly active, and three heavier mostly sterile.   The coupling of the latter with the leptons, $U_{a s}$,  is  strongly correlated with their masses 
(the naive seesaw scaling being $|U_{as}|^2 \propto M^{-1}$). The possibility that such neutrino sterile states could be responsible for any of the anomalies found in various experiments
 is of course very interesting, since it could open a new window into establishing the new physics of neutrino masses. 

Models with extra light sterile neutrinos with masses in the range of ${\mathcal O}($eV) 
could provide an explanation to the LSND/MiniBOONE \cite{Aguilar:2001ty,Aguilar-Arevalo:2013pmq} and reactor anomalies \cite{reactor}.
Sterile species in the ${\mathcal O}($keV) range could still be valid candidates for warm dark matter 
\cite{Dodelson:1993je, Shi:1998km, Abazajian:2001nj, Asaka:2005an}. The recent 
measurement of an X-ray signal \cite{Bulbul:2014sua,Boyarsky:2014jta} might be the first experimental indication of such 
possibility.  Species in the ${\mathcal O}($GeV) range could account for the baryon asymmetry in 
the Universe \cite{Akhmedov:1998qx,Asaka:2005pn} (for a recent review see \cite{Canetti:2012kh}).

There are important constraints on low-scale models from direct searches and rare processes 
such as $\mu\rightarrow e\gamma$ and $\mu e$ conversion. Recent results can be found in 
\cite{Abada:2007ux,shaposhnikov,Atre:2009rg}. The constraints are strongly dependent on 
$M$ for $M \lesssim {\mathcal O}$($100$ GeV). 

It is well known that if light sterile neutrinos with significant active-sterile mixing exist they can contribute significantly
to the energy density of the Universe. Mechanisms to reduce this contribution have been proposed, such as the presence of 
primordial lepton asymmetries~\cite{Foot:1995qk} or new interactions \cite{Hannestad:2013ana,Dasgupta:2013zpn}, which however typically require new physics beyond that of the 
sterile species. The energy density of the extra neutrino species, $\epsilon_s$, is usually quantified in terms of 
$\Delta N_{\rm eff}$ (when they are relativistic) defined by 
\begin{equation}
 \Delta N_{\rm eff} \equiv \frac{\epsilon_{s}}{\epsilon^0_\nu},
\end{equation}
where $\epsilon^0_\nu$ 
is the energy density of one SM massless neutrino with a thermal distribution 
(below $e^\pm$ annihilation it is $\epsilon^0_\nu \equiv (7 \pi^2/120) (4/11)^{4/3} T_\gamma^4$  at the photon 
temperature $T_\gamma$). One fully thermal extra sterile state that decouples from the thermal bath being relativistic 
contributes $\Delta N_{\rm eff} \simeq 1$ when it decouples. 

$N_{\rm eff}$ at big bang nucleosynthesis (BBN) strongly influences
the  primordial helium production. A recent analysis of BBN bounds \cite{Cooke:2013cba} gives $N_{\rm eff} ^{BBN} = 3.5\pm 0.2$. 
$N_{\rm eff}$ also affects the anisotropies of  the cosmic microwave background  (CMB).   
Recent CMB measurements
from Planck give  $N_{\rm eff}^{\rm CMB} = 3.30 \pm 0.27$ \cite{Ade:2013zuv}, which includes WMAP-9 polarization data \cite{WMAP} and high multipole measurements from the South Pole Telescope 
\cite{SPT} and the Atacama Cosmology Telescope \cite{ACT}. Recent global analyzes, including the 
BICEP2 results \cite{Ade:2014xna,Ade:2014gua}, seem to prefer larger values of $N^{\rm CMB}_{\rm eff}$ \cite{Giusarma:2014zza,Archidiacono:2014apa,Bergstrom:2014fqa}.

The contribution of extra sterile neutrinos to $N_{\rm eff}$  has been extensively studied in phenomenological models, where 
there is no correlation between masses and mixing angles \cite{Dolgov:2003sg}-\cite{Melchiorri:2008gq}. 
For recent analyzes of eV scale neutrinos, with and without lepton
asymmetries, see \cite{Hannestad:2012ky}-\cite{Mirizzi:2013kva}. In \cite{Hernandez:2013lza} we explored 
systematically the contribution to $N_{\rm eff}$ of the minimal Type I seesaw 
models with just two extra singlets, $N=2$. We found that whenever the two heavier states are below ${\mathcal O}(100$ MeV), 
they contribute too much energy/matter density to the Universe, while   the  possibility of having one state $\lesssim $eV and another heavier than $~100$~MeV may  not be excluded by cosmological and oscillation data constraints, but requires further scrutiny. 

The purpose of this paper is to  perform the same study in the next-to-minimal seesaw model where $N=3$. This is the standard Type I seesaw model with a low-scale, and is also often 
referred to as the $\nu$MSM. This model has been extensively studied in the literature, concentrating on 
regions of parameter space where the lightest sterile state could be a warm dark matter particle, and the two 
heavier states could be responsible for the baryon asymmetry in the Universe \cite{Asaka:2005pn}. What we add in this paper is a 
systematic study of the full parameter space to understand the constraints on the seesaw scale(s) from the modifications to the
standard cosmology induced by the three heavy neutrino states. We will assume that primordial lepton asymmetries are 
negligible. Although the model in principle satisfies the Sakharov conditions to generate a lepton asymmetry, previous works 
indicate that significant lepton asymmetries can only be generated when at least two of the sterile states are heavy enough,
$\mathcal{O}($GeV$)$, and extremely degenerate~\cite{Shaposhnikov:2008pf}.  Here we will concentrate on studying the bounds 
from cosmology when such a extreme degeneracy of the sterile neutrino states is not present. We show that, in spite of the large parameter space, the thermalization of the sterile states in 
this model is essentially controlled by one parameter: the lightest neutrino mass. 

The paper is organized as follows. In section \ref{sec:thermal} we review the  estimates of the 
thermalization rate of the sterile states as derived in \cite{Hernandez:2013lza}, which allow us to 
efficiently explore the full parameter space of the model. In section \ref{sec:bounds} we derive 
analytical bounds for the thermalization rate and in section \ref{sec:lightest} we correlate 
$\Delta N_{\rm eff}$ with the lightest neutrino mass. In section 
\ref{sec:num} we present numerical results from solving the Boltzmann equations and finally in 
section \ref{sec:bb0nu} we analyze the impact on neutrinoless double beta decay. In section
\ref{sec:conclusions} we conclude.

\section{Thermalization of sterile neutrinos in $3+3$ seesaw models} 
\label{sec:thermal}

  The model is described by the most general renormalizable Lagrangian including $N=3$ extra singlet Weyl fermions, $\nu_R^i$:
   \begin{eqnarray}
{\cal L} = {\cal L}_{SM}- \sum_{\alpha,i} \bar L^\alpha Y^{\alpha i} \tilde\Phi \nu^i_R - \sum_{i,j=1}^3 \frac{1}{2} \bar\nu^{ic}_R M_N^{ij} \nu_R^j+ h.c., \nonumber
\label{eq:lag}
\end{eqnarray}
where $Y$ is a $3\times 3$ complex matrix and $M_N$ a diagonal real matrix. The spectrum of this theory has six massive 
Majorana neutrinos, and the mixing is described in terms of six angles and six CP phases. 

We assume that the eigenvalues of $M_N$ are significantly larger than the atmospheric and solar neutrino mass splittings, which implies a hierarchy $M_N\gg Y v$ and therefore the seesaw approximation is good.  A convenient parametrization in this case is provided by that of Casas-Ibarra \cite{Casas:2001sr}, or its extension to all orders in the seesaw expansion as described in \cite{Donini:2012tt} (for an alternative see \cite{Blennow:2011vn}). 
The mass matrix can be written as
\begin{eqnarray}
{\mathcal M}_\nu = U^*~ {\rm Diag}(m_l,M_h)~ U^\dagger.
\end{eqnarray}
where $m_l= {\rm Diag}(m_1,m_2,m_3)$ and $M_h= {\rm Diag}(M_1,M_2,M_3)$. Denoting by $a$ the active/light neutrinos and $s$ the sterile/heavy species, the  unitary matrix can be written as
\begin{eqnarray}
U =  \left(\begin{array}{lll}  U_{aa} & U_{as} \\
U_{sa} & U_{ss} 
\end{array}\right), 
\label{eq:u5}
\end{eqnarray}
with
\begin{eqnarray}
U_{aa} = U_{PMNS}  {\mathcal H}, \;
U_{ss} = \overline{\mathcal H}, \;\;
U_{sa} =  i \overline{{\mathcal H}} M_h^{-1/2} R m_l^{1/2}
 , \;
 U_{as} = i U_{PMNS} {\mathcal H} m_l^{1/2} R^\dagger M_h^{-1/2} .\nonumber\\
  \label{eq:param}
\end{eqnarray}
where $U_{PMNS}$ is a $3\times 3$ unitary matrix and $R$ is a generic $3 \times 3$ orthogonal complex  matrix, while ${\mathcal H}$ and $\bar{{\mathcal H}}$ are 
defined by
\begin{eqnarray}
{\mathcal H}^{-2} = I + m_l^{1/2} R^\dagger M_h^{-1} R m_l^{1/2}, \;\;\; \nonumber\\
 \overline{{\mathcal H}}^{-2} = I + M_h^{-1/2} R m_l R^\dagger M_h^{-1/2}.
 \label{eq:hbar}
\end{eqnarray}
At leading order in the seesaw expansion, i.e. up to ${\mathcal O}\left(\frac{m_l}{M_h}\right)$, ${\mathcal H}\simeq \overline{{\mathcal H}} \simeq 1$, and we
recover the Casas-Ibarra parametrization. In this approximation $U_{PMNS}$ is the light neutrino mixing matrix measured in oscillations.

Neutrino oscillation data fix two of the three eigenvalues in $m_l$ and the three angles 
in $U_{PMNS}$. However all the heavy masses in $M_h$, the lightest neutrino mass in $m_l$, the three 
complex angles in $R$ and the three CP violating phases in $U_{PMNS}$ are presently unconstrained\footnote{There is an upper bound 
for the lightest neutrino mass but no lower bound exists at present.}. 

In \cite{simple:1990} a simple estimate for the thermalization of one sterile neutrino in the early Universe,
neglecting  primordial lepton asymmetries, was given as follows. Assuming that the active neutrinos are in thermal equilibrium with a collision rate 
given by $\Gamma_{\nu_\alpha}$, the collision rate for the sterile neutrinos can be estimated to be
\begin{eqnarray} 
\Gamma_{s_j} \simeq  \frac{1}{2} \sum_ a \langle P(\nu_a \rightarrow \nu_{s_j})\rangle \times \Gamma_{\nu_\alpha},
\label{eq:gammas}
\end{eqnarray}
where $\langle P(\nu_\alpha \rightarrow \nu_{s_j})\rangle$ is the time-averaged  probability 
$\nu_\alpha \rightarrow \nu_{s_j}$. This probability depends strongly on temperature
because the neutrino index of refraction in the early Universe is modified by coherent scattering of neutrinos with 
the particles in the plasma \cite{Notzold:1987ik}. Thermalization will be achieved if there is any temperature where this 
rate is higher than the Hubble expansion rate, i.e. $\Gamma_{s_j}(T) \geq H(T)$. In a radiation-dominated 
Universe, $H(T) = \sqrt{\frac{4 \pi^3 g_*(T)}{45}} \frac{T^2}{M_{\rm Planck}}$, with $g_*(T)$ the number of relativistic degrees of freedom.

  One can therefore
define the function $f_{s_j}(T)$, which measures the sterile production rate of the species $s_j$ in units of the Hubble expansion rate, 
\begin{equation}
f_{s_j}(T)\equiv \frac{\Gamma_{s_j}(T)}{H(T)}.
\end{equation}
It reaches a maximum at  some temperature, $T_{\rm max}$ \cite{simple:1990}. If $f_{s_j}(T_{\rm max}) \geq 1$, the sterile state will reach a thermal abundance at early times.  We can    
 estimate the contribution to $N_{\rm eff}$ as 
\begin{equation}
N_{\rm eff} \simeq N_{\rm eff}^{SM}+\sum_j \left(1-\exp(-\alpha f_{s_j}(T^j_{\rm max}))\right),
\end{equation}
at decoupling if they are still relativistic, where $\alpha$ is an ${\mathcal O}(1)$ numerical constant. {Provided $f_{s_j}(T^j_{\rm max})$ is sufficiently larger than one, $N_{\rm eff}$ saturates to the number of thermalized species, up to exponentially small corrections.}

In \cite{Hernandez:2013lza}, this result was also derived from the Boltzmann equations \cite{oldies,polariz, 
Sigl:1992fn,Dolgov:2002wy}, in the assumption of no primordial large lepton asymmetries.
As shown in Appendix A \ref{sec:appendix}, in spite of the complex $6\times 6$ mixing, the thermalization of 
the sterile state $j$ is roughly given by the sum of three  $2\times 2$ mixing contributions in agreement with the naive 
expectation of eq.~(\ref{eq:gammas})
\be
f_{s_j}\left(T\right) = \sum_{\alpha=e,\mu,\tau} \frac{\Gamma_{\nu_\alpha}\left(T\right)}{H\left(T\right)} 
\left(\frac{M^2_{j}}{2 p V_\alpha(T)-M^2_{j}}\right)^2
|\left(U_{a s}\right)_{\alpha j}|^2,
\ee
where $p$ is the momentum, $V_\alpha(T)$ is the potential induced by coherent scattering in the plasma \cite{Notzold:1987ik}
and $\Gamma_{\nu_\alpha}(T)$ is the scattering 
rate of the active neutrinos.  Both $V_\alpha$ and $\Gamma_\alpha$ depend on the temperature since the number 
of scatters increases with $T$ \cite{Abazajian:2001nj,Abazajian:2002yz,Asaka:2006nq}. 
While the former varies only when the lepton states become populated, the latter depends significantly 
on the quark degrees of freedom and therefore changes significantly at the QCD phase transition. The quark 
contribution to $\Gamma_{\nu_\alpha}$ is however rather uncertain, we therefore neglect this contribution, since this is a conservative
assumption if we want to minimize thermalization: any contribution that will increase $\Gamma_{\nu_\alpha}$ would help 
increase the thermalization rate.
 
 The most complete calculation of $\Gamma_{\nu_\alpha}$ has been presented in \cite{Asaka:2006nq}, where 
 a full two-loop computation of the imaginary part of the neutrino self-energy was presented. The results for the leptonic contribution to 
 $\Gamma_{\nu_\alpha}(T)$ can be accurately parametrized by in terms of $C_\alpha(T)$ as
 \be
 \Gamma_{\nu_\alpha} \simeq {C_\alpha(T)} G_F^2 T^4 p
 \ee
 that can be extracted from the numerical results of \cite{Asaka:2006nq}, recently made publicly available in 
 ref.~\cite{laine}.

For temperatures above the different lepton thresholds, the results can be approximated by

\begin{itemize}

\item[($\tau$)] $T\gtrsim180$ MeV: $C_{e,\mu,\tau} \simeq 3.43$ and $V_\alpha= A\,T^4p$ for $\alpha=e,\mu,\tau$;

\item[($\mu$)] $20$ MeV $\lesssim T\lesssim 180$ MeV: $C_{e,\mu} \simeq 2.65$, $C_{\tau} \simeq 1.26$,
$V_e=V_\mu= A\,T^4p$ and  $V_\tau= B\,T^4p$;

\item[($e$)] $T\lesssim 20$ MeV: $C_e \simeq 1.72$, $C_{\mu,\tau} \simeq 0.95$, 
$V_e=A\,T^4p$ and  $V_\mu= V_\tau= B\,T^4p$,
\end{itemize}
with 
\begin{eqnarray}
B \equiv-2\sqrt{2}\,\left(\frac{7\zeta(4)}{\pi^2}\right)\frac{G_F}{M_Z^2},\;\;\; A\equiv B -4\sqrt{2}\,  \left(\frac{7\zeta(4)}{\pi^2}\right)\frac{G_F}{M_W^2}.
\end{eqnarray}

\begin{figure}
\begin{center}
\includegraphics[scale=0.8]{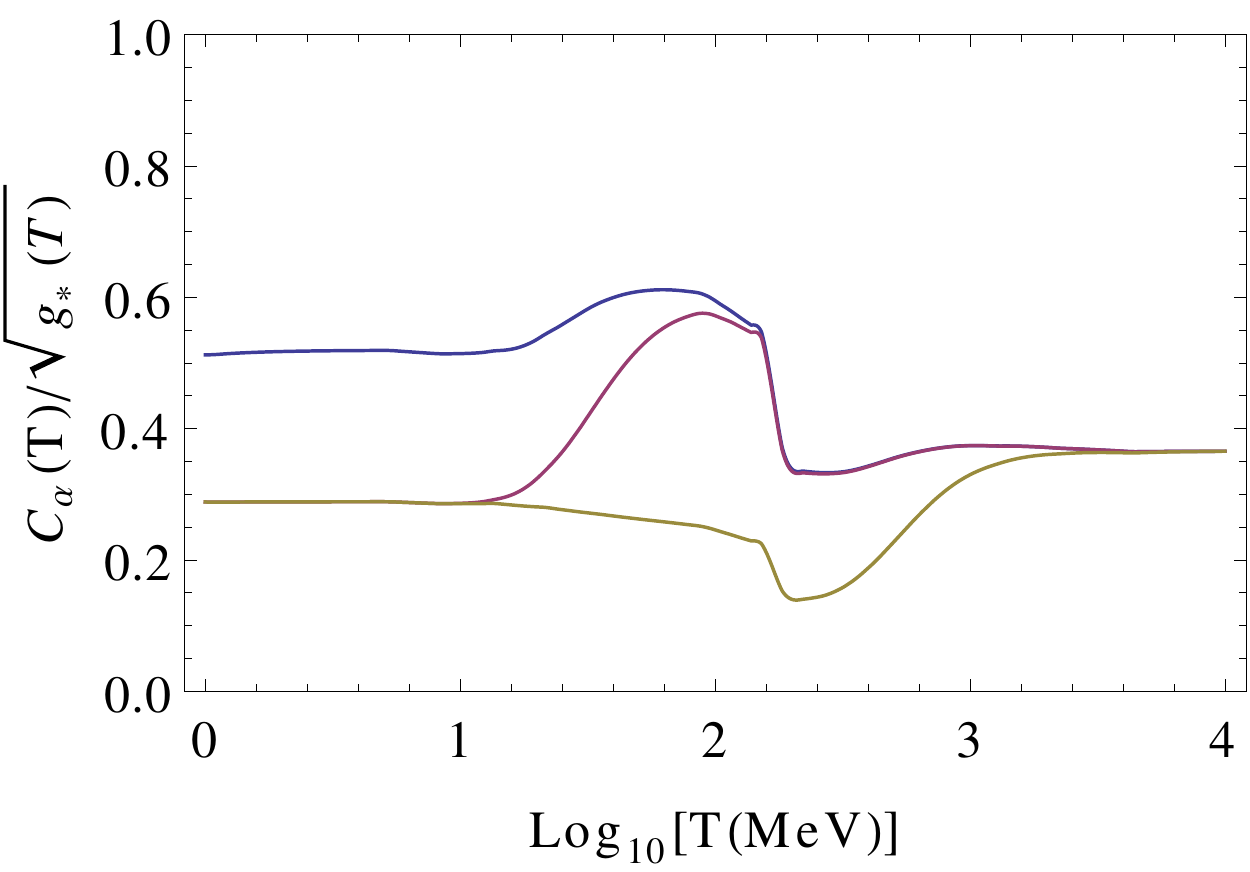}
\caption{\label{fig:Calpha} Leptonic contribution to ${C_\alpha(T)}/\sqrt{g_{*}(T)}$ taken from 
refs.~\cite{Asaka:2006nq,laine} for $\alpha=e\,{\rm (top/blue),\; }\mu\,{\rm (middle/magenta),\; }\tau\,{\rm (bottom/yellow) }$ .}
\end{center}
\end{figure}

In Fig.~\ref{fig:Calpha} we show
${C_\alpha(T)}/\sqrt{g_{*}(T)}$ as a function of the temperature. We include the $T$ dependent 
normalization factor, $\sqrt{g_{*}(T)}$, coming from $H(T)$. Note that the dependence on the temperature 
of this factor is small.

Let $T_{\rm max}$ be the value of the temperature at which $f_{s_j}\left(T\right)$ is maximum 
\footnote{We note that $T_{\rm max}$ depends on $M_j$, but to simplify notation we skip the index $j$ in 
this quantity.}. For  $p=3.15\,T$, and neglecting the $T$ dependence of $C_\alpha/\sqrt{g_{*}}$, 
$T_{\rm max}$ is bounded by
\be
T^\tau_{\rm max}\equiv\left(\frac{M_j^2}{59.5~ |A|}\right)^{1/6} \leq T_{\rm max} \leq \left(\frac{M_j^2}{59.5~ |B|}\right)^{1/6} .
\label{eq:tmax}
\ee
Thermalization will take place provided $f_{s_j}(T_{\rm max}) \geq 1$. In the next section we derive an analytical lower bound on this quantity, which can 
be translated therefore into a sufficient condition for thermalization. 

\section{Analytical bounds}
\label{sec:bounds}

For a given set of mixing and mass parameters we have the following 
general lower bound for $f_{s_j}(T)$:
\be
\label{fss}
f_{\rm B}\left(T\right)\equiv {\rm Min}\left[\frac{C_\tau(T)}{ \sqrt{g_{*}(T)} }\right] \frac{G_F^2 pT^4 \sqrt{g_{*}(T)}}{H(T)}
\left(\frac{M^2_{j}}{2 p V_e-M^2_{j}}\right)^2\sum_{\alpha=e,\mu,\tau}|\left(U_{a s}\right)_{\alpha j}|^2
\leq f_{s_j}\left(T\right).
\ee 
This results from the fact that $|V_e| \geq |V_\alpha|$ and $C_\alpha \geq C_\tau$ for all $\alpha=e,\mu,\tau$. The minimization of
$C_\tau/\sqrt{g_*}$ as function of $T$, gets rid of the $T$ dependence of this factor. 

The function $f_{\rm B}(T)$ is maximized at $T^\tau_{ max}$, defined in eq.~(\ref{eq:tmax}). It then follows that
\be
f_{\rm B}\left(T_{max}^\tau\right) \leq  f_{s_j}\left(T^\tau_{max}\right)
\leq f_{s_j}(T_{max}).
\ee
In summary, taking the average momentum, $p=3.15T$, $f_{s_j}\left(T_{\rm max}\right)$ is bounded by
\be
f_{s_j}\left(T_{\rm max}\right) \geq f_{\rm B}(T^\tau_{\rm max}) = \frac{\sum_{\alpha}|\left(U_{as}\right)_{\alpha j}|^2 M_j}{3.25 \cdot 10^{-3} {\rm eV}}.
\label{eq:bound}
\ee
Using eq.~(\ref{eq:param}) in the Casas-Ibarra limit, the dependence on the parameters of the model 
in the above equation can be simplified to the following combination:
\bea
\sum_{\alpha}|\left(U_{as}\right)_{\alpha j}|^2 M_j &=&
\sum_\alpha (U_{PMNS} m_l^{1/2}R)_{\alpha j}(R^\dagger m_l^{1/2}U_{PMNS}^\dagger)_{j\alpha} 
=\left(R^\dagger m_l R \right)_{jj}\equiv h_j.\nonumber\\
\label{eq:h_j}
\eea
Therefore the analytical lower bound does not depend on the angles and CP-phases of the PMNS matrix. It depends 
only on the undetermined Casas-Ibarra parameters and the light neutrino masses. The lower bound can be further 
simplified using
\bea
h_j = \sum_\alpha |R_{\alpha j}|^2 m_\alpha \geq |\sum_\alpha R^2_{\alpha j} m_\alpha|\geq \lvert\sum_\alpha  R_{\alpha j}^2 m_1\lvert = m_1,
\label{eq:m1}
\eea
where in the last step we have used the orthogonality of the $R$ matrix and assumed a normal hierarchy of the light neutrinos 
(NH). The result for an inverted hierarchy (IH) is the same substituting $m_1 \rightarrow m_3$. Finally using 
Eqs.~(\ref{eq:h_j}) and (\ref{eq:m1}) in eq.~(\ref{eq:bound}) we obtain 
\be
f_{s_j}\left(T_{\rm max}\right) \geq 
\frac{h_j}{3.25\cdot 10^{-3} {\rm eV}}\geq 
\frac{m_1}{3.25\cdot 10^{-3} {\rm eV}}\equiv\frac{m_1}{m_1^{th}},
\label{eq:g}
\ee
which defines $m_1^{th}$.

\section{Lightest neutrino mass versus thermalization}
\label{sec:lightest}

The  thermalization  of $j$-th heavy sterile state will occur provided $f_{s_j}(T) \geq 1$ for some $T$. Therefore a 
sufficient condition is that $f_{s_j}(T_{\max}) \geq 1$ or using eq.~(\ref{eq:g}) $m_1 \geq m_1^{th}$. From the analytical 
bound we therefore deduce that thermalization of the three states will occur if 
\bea
m_1 \geq  3.25\cdot 10^{-3} {\rm eV} ,
\label{eq:m1_bound}
\eea
for any value of the unconstrained parameters in $R$ and the CP phases.  
We note that a more restrictive upper bound on the lightest neutrino mass was derived in \cite{Asaka:2005an,Asaka:2006nq} 
under the assumption that $M_1$ was a warm dark matter candidate in the keV range. 

\begin{figure}
\begin{center}
\includegraphics[scale=0.9]{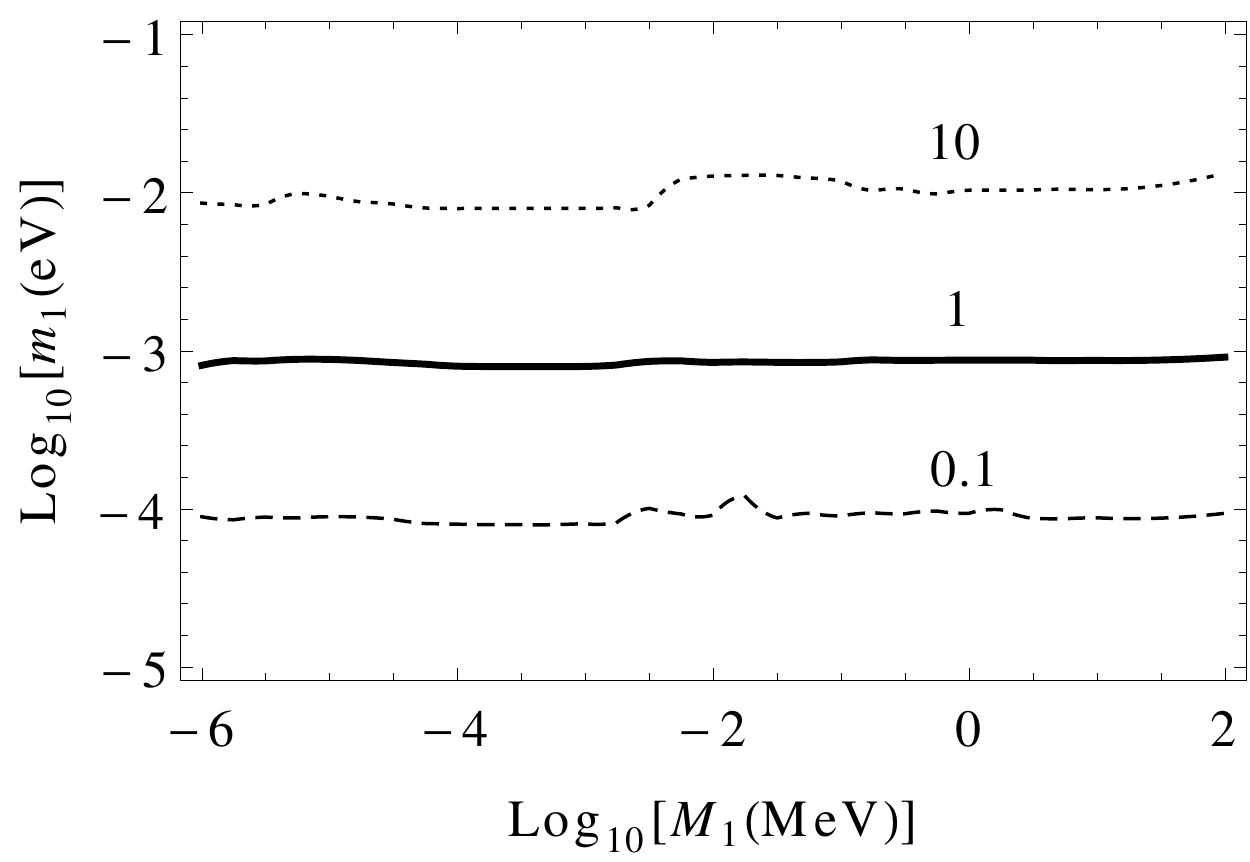}
\caption{\label{fig:fs1} Contours of Min[$f_{s_1}(T_{\rm max})$]=0.1, 1, 10 on the plane $(M_1,m_1)$.}
\end{center}
\end{figure}

In Fig.~\ref{fig:fs1} we show the contour plots of the minimum of $f_{s_1}({T_{\rm max}})$ (varying  the unconstrained parameters in $R$ and the CP phases in the full  range), as a function of $m_1$ and 
$M_1$. The three lines correspond to ${\rm Min}[f_{s_1}(T_{\rm max})]=10^{-1}, 1, 10$. As expected the minimum 
is strongly correlated with $m_1$ and is roughly independent of $M_1$. Values of $m_1$ below the contour line at 1 correspond to
non-thermalization, therefore we read 
\be
m_1 \leq {\mathcal O}(10^{-3} {\rm eV}), 
\ee
for $M_1 \in [1$eV-$100$MeV$]$. The numerical bound is slightly stronger than the analytical bound given by eq.~(\ref{eq:m1_bound}). Had we 
considered any other of the heavy states $j=2,3$ the results would be the same (i.e. the same minimum of $f_{s_j}(T_{\rm max})$ 
would be obtained for different values of the unconstrained parameters).

\begin{figure}
 \begin{center}
\includegraphics[scale=1.2]{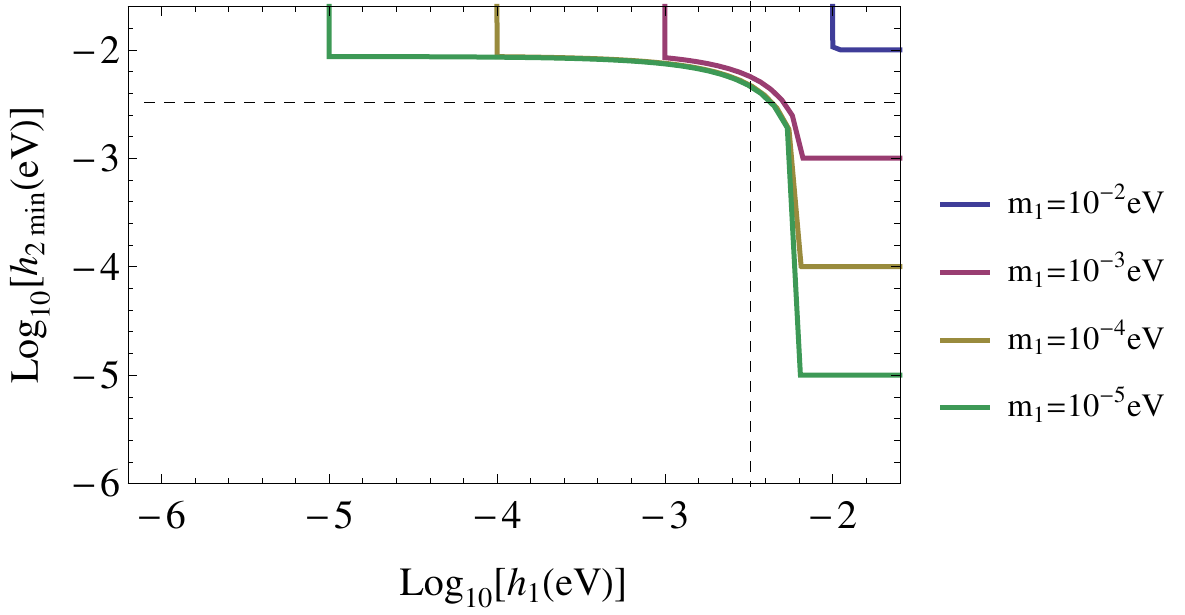}
\caption{\label{fig:hvsh} Minimum of $h_2$ in bins of $h_1$ in the full allowed parameter space with fixed 
$m_1= 10^{-[5-2]} $eV. The dashed line corresponds to the analytical bound $m_1^{th}=3.25\cdot 10^{-3} {\rm eV}$.}
\end{center}
\end{figure}

A less stringent (sufficient) condition for thermalization of the state $j$ is 
\be
h_j\geq m_1^{th}
\label{eq:h_bound}
\ee
as it follows from eq.~(\ref{eq:g}). It turns out that this condition is always satisfied for {\it at least two of the three heavy neutrinos}, independently of $m_1$ or the Casas-Ibarra parameters.  In Fig.~\ref{fig:hvsh} we show the minimization of $h_2$ in the 
full parameter space within each bin of $h_1$, shown in the $x$-axis, for fixed values of $m_1$. 
Although either $h_1$ or $h_2$ can always be below the
$m_1^{th}$ line (shown as dashed line) if $m_1 \leq m_1^{th}$, the other one is always significantly above it. The same pattern is observed with any 
pair of $h_j$. This shows that at most one of the sterile states might not thermalize, and to have one not thermal 
requires that $m_1 \leq m_1^{th}$.

It is easy to see how $h_j$ can reach its lower bound, $m_1$,   without contradicting present  neutrino data. One can
always choose $R_{\alpha j} =0$ for $\alpha \neq j$. For $j=1$,  the orthogonal matrix reduces to the form 
  \begin{eqnarray}
  R = \left(\begin{array}{ll} 1 & 0  \\
  0 & R_{2\times 2} \end{array}\right),
  \end{eqnarray}
where $R_{2\times 2}$ is an orthogonal two-dimensional matrix that depends on one complex angle. 
For $j=2,3$ the matrix is analogous with the appropriate permutation of the heavy states. The model 
therefore reduces in this limit to a $3+2+1$, where one sterile state is essentially decoupled. When $m_1 \leq m_1^{\rm th}$, the latter 
might thermalize or not depending on the unknown parameters, while the other two states always thermalize, as in 
 the minimal $3+2$ model already considered in ref.~\cite{Hernandez:2013lza}. 

 In the next section we evaluate the implications for 
$N_{\rm eff}$ in both cases.
     
\section{$N_{\rm eff}$ in the $3+3$ model}
\label{sec:num}

\subsection{$m_1 \geq m_1^{\rm th}$}

In this case, the three sterile states thermalize, each of them contributing with 
$\Delta N^{(j)}_{\rm eff}(T_{d_j}) \approx1$ at their decoupling temperature, $T_{d_j}$ (provided they are still
relativistic). This contribution gets diluted later on, due to the change of $g_*(T)$ between $T_{d_j}$ and the active 
neutrino decoupling, $T_{BBN}$, when BBN starts.  The dilution factor is relevant only for masses larger than 
$M_j \gtrsim $1~keV \cite{Hernandez:2013lza}. 

If they are still relativistic at $T_W$, we can therefore estimate
\bea
\Delta N_{\rm eff}^{BBN} = \sum_j \left( \frac{g_*(T_{BBN})}{g_*(T_{d_j})}\right)^{4/3},
\label{eq:dneff}
\eea
where the sum runs over the three heavier states. 

For $M_j \geq {\mathcal O}(100)$ MeV, the contribution to the energy density could  be significantly 
suppressed with respect to the estimate eq.~(\ref{eq:dneff}), because either they decay sufficiently 
early before BBN and/or become non-relativistic at $T_{d_j}$ and therefore get Boltzmann suppressed. Additional constraints will be at work in some regions of parameter space
even for those larger masses, but they are likely to depend on the unknown mixing parameters, so we concentrate on the case where at least one of the three heavy neutrinos has a  mass below this limit. 

We consider in turn the following possibilities.

\begin{itemize}
\item For all $j$, $M_j \lesssim$ 100 MeV
\end{itemize}

After recent measurements, the BBN constraints mentioned in the introduction give  $\Delta N_{eff}^{BBN} \leq 0.9$ at 2$\sigma$.
From the results of \cite{Hernandez:2013lza} in the 3+2 model,  we estimate that 
$M_j \lesssim 10-100$keV would be excluded from BBN bounds in this case. For larger masses, dilution is sufficiently 
strong to avoid BBN bounds, but the contribution to the energy density after BBN is anyway too large. When they become 
non-relativistic, their contribution to the energy density can be estimated to be \cite{KolbTurner}
\begin{eqnarray}
\Omega_{s_j} h^2=  10^{-2} M_j(eV) \Delta N_{eff}^{(j)BBN},
\end{eqnarray}
where $\Delta N_{eff}^{(j) BBN}$ is estimated from the ratio of number densities of the $j$-th state and one standard neutrino at BBN. 
  If they do not decay before recombination, 
Planck constraint  on $\Omega_m h^2$ would completely exclude, such high masses. On the other hand, if they decay, they transfer this energy density to radiation. The case 
in which they decay at  BBN or before (only for masses above 10MeV or so) has been considered in detail in \cite{Dolgov:2000pj, Ruchayskiy:2012si} and essentially BBN constraints,  combined with direct search constraints \cite{shaposhnikov,Atre:2009rg,Ruchayskiy:2011aa}, exclude the range $10-140$MeV.    If they decay after BBN, they transfer the energy density mostly to the already decoupled light neutrinos, a contribution that can be parametrized in terms of 
 $\Delta N_{\rm eff}$ which is enhanced with respect to that at BBN, eq.~(\ref{eq:dneff}), by a factor $\propto \frac{M_j}{T_{dec}^{(j)}}$ , where $T_{dec}^{(j)}$ is the decay temperature of the $j$-th species. This temperature can be estimated by the relation $H(T_{dec}^{(j)})= \tau^{-1}_{s_j}$, where 
 \begin{eqnarray}
 \tau^{-1}_{s_j} \simeq \frac{G_F^2 M^5_j}{192 \pi^3} \sum_\alpha |(U_{as})_{\alpha j}|^2,
 \end{eqnarray}
 (for $M_j$  below any lepton or hadron threshold). We are not aware of a self-consistent global 
 cosmological analysis of such scenario. Assuming that CMB constraints on extra radiation $\Delta N_{\rm eff}$ roughly apply to it, the large 
 mass region,  still allowed by BBN due to dilution, is anyway excluded by CMB measurements, because 
 the  ratio $M_j/T^{(j)}_{dec}$  is very large. Recent analyzes on dark radiation
 from decays can be found in \cite{GonzalezGarcia:2012yq,Hasenkamp:2012ii,Hasenkamp:2014hma}.

\begin{itemize}
\item $M_1,M_2 \lesssim 100$ MeV$\ll  M_3$
\end{itemize}

In this case, the results of the 3+2 model apply directly and the conclusion is the same as before: BBN constraints force the masses to be large to enhance dilution, but such heavy states contribute 
too much energy density either in the form of matter or extra radiation. 

\begin{itemize}
\item $M_1 \lesssim 100$ MeV$\ll  M_2, M_3$
\end{itemize}

In this case, any value of $M_1$ could be barely compatible with BBN constraints, since $\Delta N_{\rm eff} \leq 1$. CMB constraints would however force the state to be very light, 
sub-eV, which implies $\Delta N_{\rm eff} \simeq 1$ and therefore some tension with BBN. On the other hand, constraints from oscillations are important in this range \cite{Donini:2011jh}.

\begin{figure}
 \begin{center}
\includegraphics[scale=0.7]{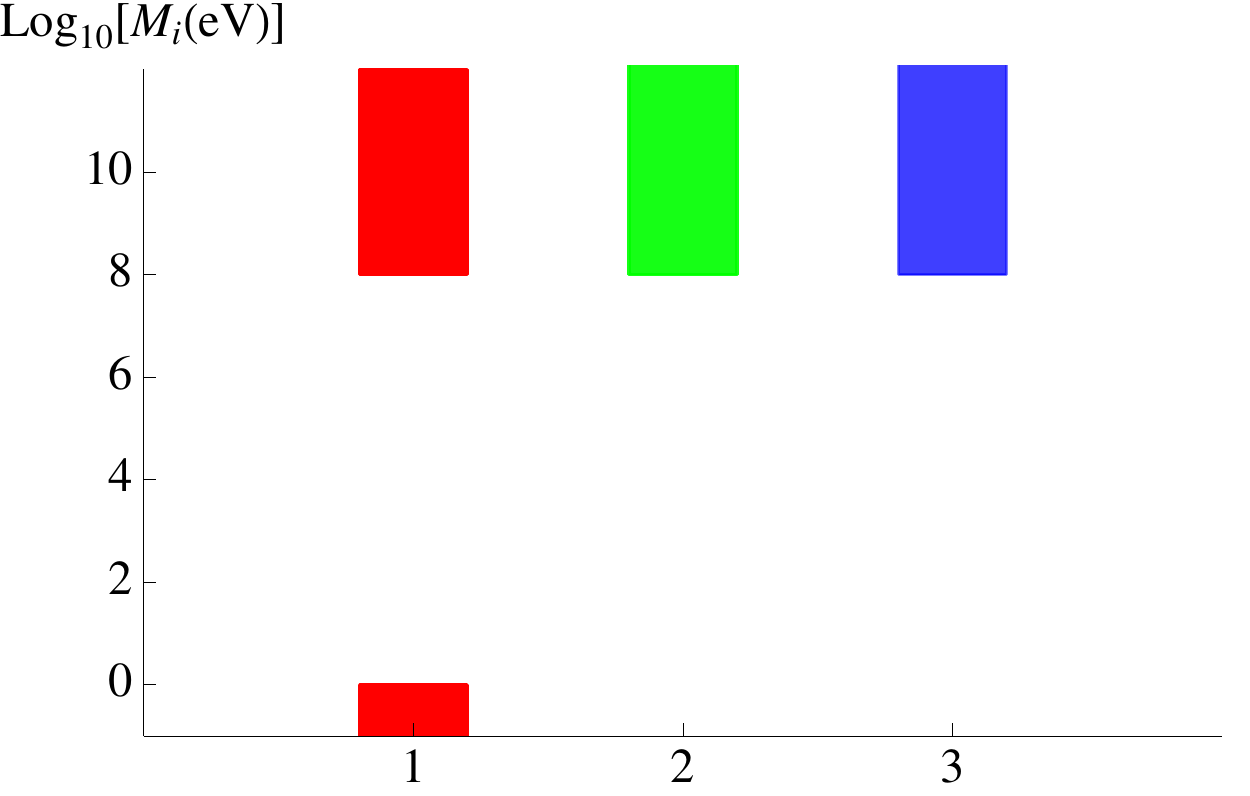}
\caption{\label{fig:gtm1th} Allowed spectra of the heavy states $M_i$ for $m_1 \geq m_1^{th}$.}
\end{center}
\end{figure}

The allowed ranges of the $M_j$ are qualitatively depicted in Fig.~\ref{fig:gtm1th}. 

\subsection{$m_1 \leq m_1^{\rm th}$}

If the lightest neutrino mass is below $m_1^{th}$, one of the states might not thermalize \footnote{There are always ranges of parameters where it does thermalize and in this case the same conclusions apply as in the previous section.}, we will take it to be the lightest sterile state although it could be any other. As shown above, this can happen in a region of parameter space with effective
decoupling of the first state. 
A more precise estimate of $\Delta N_{\rm eff}^{BBN}$ is given 
from solving the Boltzmann equations reviewed in Appendix A. We consider two cases:
\begin{itemize}
\item  The unknown mixing parameters (i.e. the Casas-Ibarra parameter of the matrix $R$ and the CP phases) are fixed by minimizing $f_{s_1}(T_{\rm max})$ and $f_{s_2}(T_{\rm max})$ as function of $m_1$ and $M_1$, and for fixed values of $M_2$ and $M_3$. 
\item The unknown parameters correspond to those that satisfy $f_{s_1}(T_{\rm max}) = 10 {\rm Min}[f_{s_1}(T_{\rm max})]$ (i.e. 
the lightest sterile state does not thermalize, but the thermalization rate is 10 times larger than its 
minimum) and minimise $f_{s_2}(T_{\rm max})$.
\end{itemize}

\begin{figure}
 \begin{center}
\includegraphics[scale=0.7]{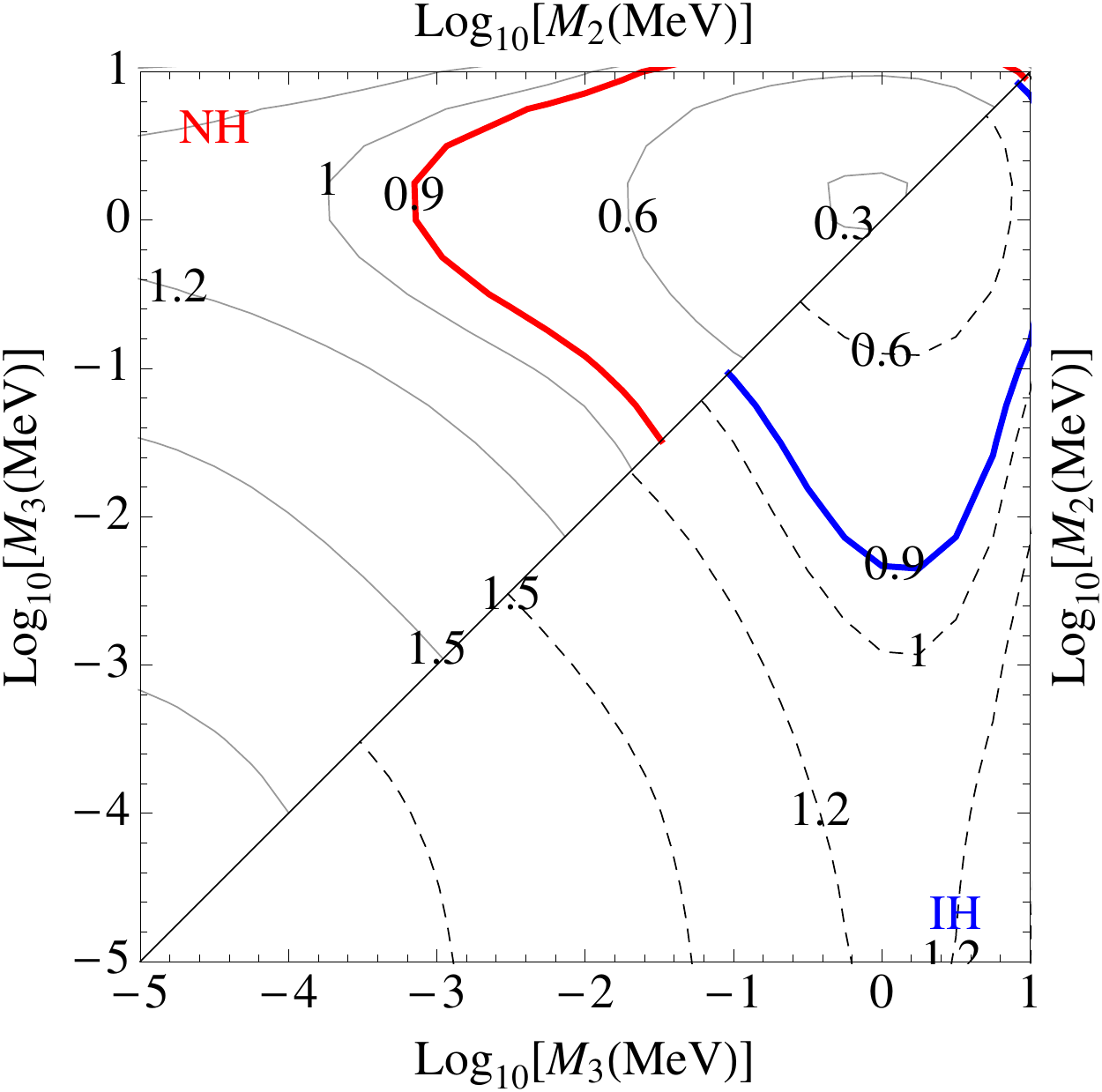}
\caption{\label{fig:dneff23} $\sum_{j=2,3} \Delta N_{\rm eff}^{(j)BBN}$ for $m_1\leq m^{th}_1$, as a function of $M_2$ and $M_3$. The thick lines correspond 
to present BBN bounds. }
\end{center}
\end{figure}

\begin{figure}
 \begin{center}
\includegraphics[scale=0.7]{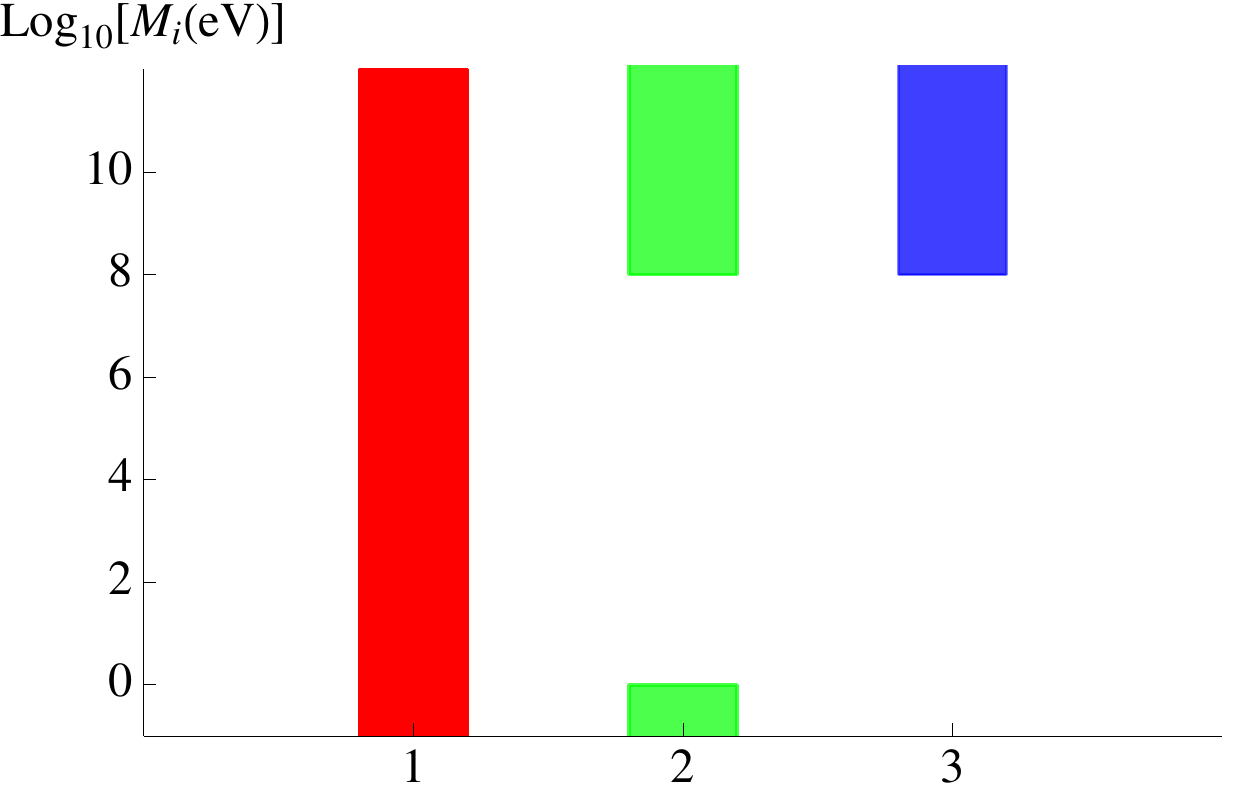}
\caption{\label{fig:ltm1th} Allowed spectra of the heavy states $M_i$ for $m_1 \leq m_1^{th}$. The unconstrained mass could be any $i=1, 2,3$. }
\end{center}
\end{figure}

In Fig.~\ref{fig:dneff23} we show the  contribution  $\sum_{j=2,3} \Delta N_{\rm eff}^{(j)BBN}$ for the NH(IH) cases. It is approximately the same as that found in 
the $3+2$ model \footnote{The small differences with respect to the results in \cite{Hernandez:2013lza} are due to the more precise scattering rates $\Gamma_\alpha$ used here. } and  independent of $m_1$ and $M_1$ .  On the other hand, the contribution 
$\Delta N_{\rm eff}^{(1)BBN}$  depends strongly on $m_1$ and it is roughly 10 times larger in the second case than in the first,
as expected from Fig.~\ref{fig:fs1}. Assuming that the contribution of the non-thermal state is negligible, the model is still 
strongly disfavored if $M_2, M_3 \lesssim 100$MeV, as explained above. The case with $M_2 \lesssim 100$MeV $\ll M_3$ could be 
barely compatible with BBN and CMB constraints if $M_2\lesssim$eV. The allowed ranges of the $M_j$ are qualitatively depicted 
in Fig.~\ref{fig:ltm1th}.

\begin{figure}
  \begin{center}
\includegraphics[scale=0.8]{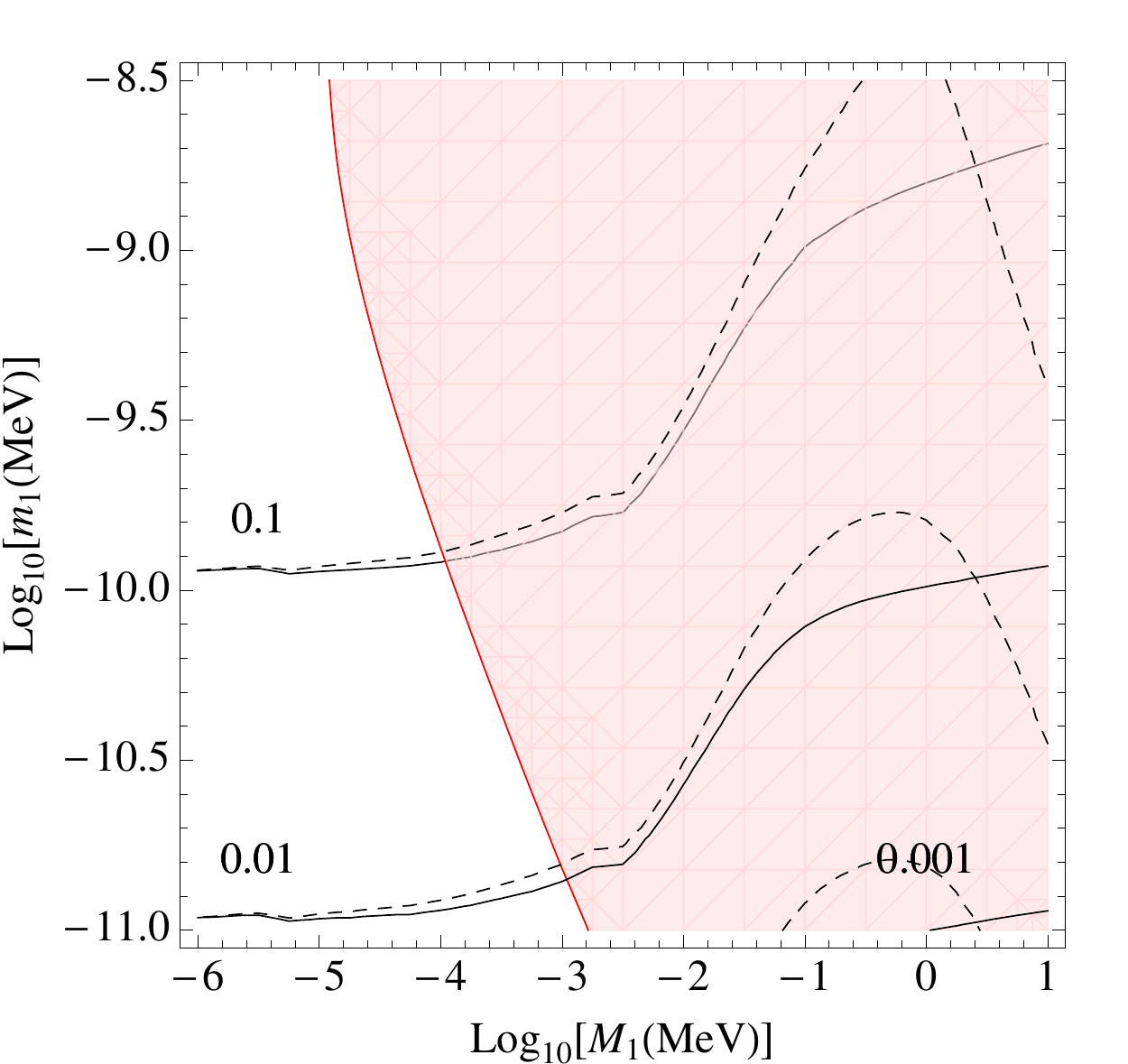}
\caption{\label{fig:non thermal} Contour plots for $\Delta N^{(1)BBN}_{eff}= 10^{-1},10^{-2},10^{-3}$ defined by the ratio 
of the energy density of the $j=1$ sterile state and one standard neutrino as a function of $m_1$ and $M_1$. The solid (dashed)
lines correspond to the contours of the ratio of sterile to active number (energy) densities. The shaded region corresponds 
to $\Omega_{s_1} h^2 \geq 0.1199$ and the dashed straight line is roughly the one corresponding to decay at recombination. 
The heavier neutrino masses
have been fixed to $M_{2,3}$=1GeV, 10GeV and the unconstrained parameters have been chosen to minimise $f_1(T_{\rm max})$ and $f_2(T_{\rm max})$. The light neutrino spectrum has been assumed to be normal (NH).  }
\end{center}
\end{figure}

When $M_2, M_3$ are above  $100$MeV, the only contribution to $\Delta N_{\rm eff}$ would be that of the lighter state. In
Fig.~\ref{fig:non thermal} we show the contour levels for $\Delta N_{\rm eff}^{(1) BBN}$ as obtained from the Boltzmann 
equations from the ratio of energy (number) densities of the $j=1$ sterile state and one standard neutrino at BBN (see eqs.~(\ref{eq:dneffen}) and (\ref{eq:dneffnum}) in Appendix A),   versus $m_1$ and $M_1$, assuming no lepton asymmetries.
In the case of degenerate heavier states significant lepton asymmetries can be produced \cite{Laine:2008pg}, which can modify 
significantly the production of the lighter state \cite{Foot:1996qc,Bell:1998ds,Laine:2008pg}. We will explore systematically that region of parameter space  in a future work, but here we consider only the
non-degenerate case where asymmetries are not expected to be of relevance.

In the figure we also included the line, enclosing the shaded region, corresponding to 
$\Omega_{s_1} h^2 = \Omega_m h^2 = 0.1199$,  which is the result from the PLANCK collaboration in a 
$\Lambda$CDM model \cite{Ade:2013zuv}. In the shaded region the sterile state contributes too much to the 
matter density and therefore is excluded. Further constraints from Lyman-$\alpha$ and X-rays can be found in the recent review
\cite{Canetti:2012kh}, and based on Pauli exclusion principle and Liouville's theorem in \cite{Boyarsky:2008ju}. The 
almost vertical dashed line corresponds to decay roughly at recombination, which means that in the region to the right of this curve, 
the $j=1$ state decays before, and contributes as extra radiation, roughly $\Delta N_{\rm eff}^{(1)BBN} \times\frac{M_1}{T^{(1)}_{dec}}$, which is much larger than one in the whole plane and is therefore excluded. 

We note that for $M_1$ in the keV range, where it could be a WDM candidate, the allowed region requires $m_1 \lesssim {\mathcal O}(10^{-5}$ eV), which is in good agreement with  the bound derived in \cite{Asaka:2005pn}. 

We have also studied the case where it is the $j=2$ state that does not reach thermalization, with $M_1=0.5$ eV, $M_3=1$ GeV. The contribution of the $j=2$ state, $\Delta N_{\rm eff}^{(2)}$ is essentially the same as that shown in 
Fig.~\ref{fig:non thermal}. In this case the contribution of the lighter state is $\Delta N_{\rm eff}^{(1)BBN} \simeq 1$, because dilution is very small for such light masses. 
 
All the results we have shown are for a normal hierarchy of the light neutrino spectrum, but the results for IH are almost identical if we exchange $m_1 \rightarrow m_3$.

\section{Impact on Neutrinoless Double Beta Decay}
\label{sec:bb0nu}

In the $3+3$ seesaw models studied here the light and heavy neutrinos are Majorana particles and, therefore, 
they can contribute to lepton number violating processes such as the neutrinoless double beta ($\beta\beta0\nu$)
decay. The spectra of Fig.~\ref{fig:ltm1th}, allowed if $m_1 \leq m_1^{th}$, 
will have important implications for this observable for two reasons: 1) the contribution of the light neutrinos to   the amplitude of this process, $m_{\beta\beta}$,
depends strongly on the lightest neutrino mass, 2)  sterile states with masses below 100~MeV could also contribute 
significantly to this amplitude. The contribution of states with masses well above 100~MeV would be generically subleading~\cite{Blennow:2010th,LopezPavon:2012zg}. 

\begin{figure}
\begin{center}
\includegraphics[scale=0.8]{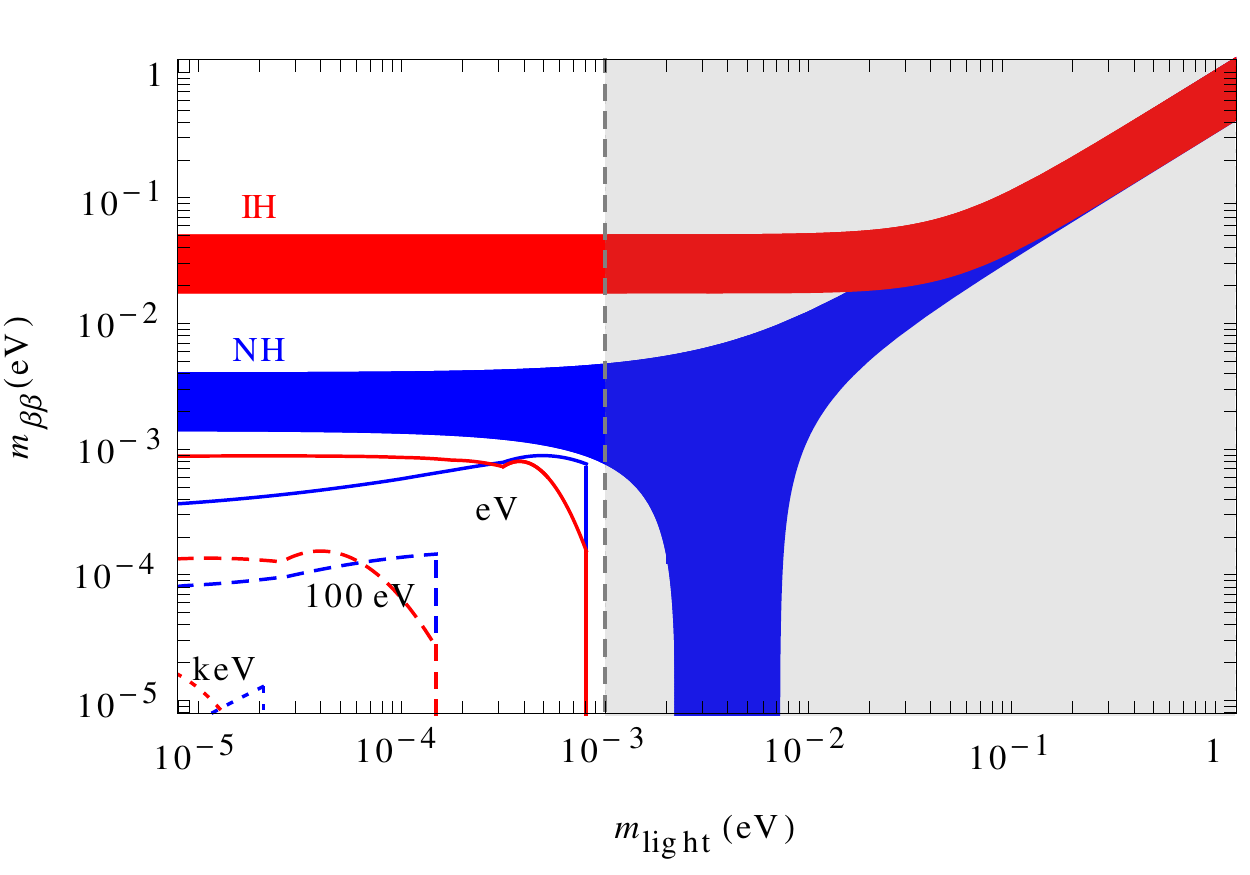}
\caption{\label{fig:mbb} 
 $m_{\beta\beta}$ as a function of the lightest neutrino mass: contribution from the active neutrinos (red and blue regions) and the maximum contribution of the lightest sterile 
neutrino, for $M_1=1$ eV (solid), 100 eV (dashed), 1 keV (dotted), for NH (blue) and IH (red) restricting 
$\Omega_{s_1} h^2 \leq 0.12$ and $f_{s_1}(T_{\rm max})\leq 1$, for $M_{2,3}\gg100$ MeV, as a function of the lightest neutrino 
mass. The shaded region is ruled out for $M_1 \in [1$eV-$100$MeV$]$ by the thermalization bound on the lightest neutrino mass, 
$m_1\leq 10^{-3}$eV.}
\end{center}
\end{figure}

If the three heavy states are well above $100$MeV, $m_{\beta\beta}$ is the standard result for the three light Majorana neutrinos. 
It is shown by the well-known  colored bands on Fig.~\ref{fig:mbb} as a function of the lightest neutrino mass, for the two 
neutrino hierarchies. If one of the states, for example $j=1$, is  in the range [1eV, 100 MeV], we have seen that it cannot have
the thermal abundance which requires 
an upper bound on the lightest neutrino, $m_1 \leq 10^{-3}$eV, shown by  the vertical dashed grey line. In this case, 
the sterile state can give a relevant contribution to the amplitude of the process 
and $m_{\beta\beta}$ reads:
\be
\label{eq:mbb}
m_{\beta\beta}= 
e^{i \alpha} m_1 c_{12}^2c_{13}^2 +  e^{i \beta} m_2 c_{13}^2 s_{12}^2 + m_3 s_{13}^2 +
\left(U_{as}\right)_{e4}^2 M_1.
\ee
The maximum value of the extra term (with the constraints that the corresponding sterile state does not thermalize, 
i.e. $f_{s_1}(T_{\rm max})\leq 1$, and it does not contribute too much to the energy density, $\Omega_{s_1} h^2 \leq 0.12$)
is shown by the  lines for $M_1$=1 eV, 100 eV and 1 keV, as function of the lightest neutrino mass, $m_{\rm light}= m_1 (m_3)$ for NH (IH).

Fig.~\ref{fig:mbb} shows that  the quasi-degenerate light neutrino spectrum is ruled out for 
$M_1 \in [1$eV-$100$MeV$]$ and $M_{2,3}\gg 100$ MeV. The region of the parameter space in which a cancellation 
can occur in the active neutrino contribution is also excluded. It is remarkable that the thermalization bound on $m_{light}$ is around two orders of 
magnitude stronger than the present constraint on the absolute neutrino mass scale from 
Planck~\cite{Ade:2013zuv}. On the other hand, we can also conclude that the contribution of the lightest 
sterile neutrino to the process is subleading and well below the (optimistic) sensitivity of the next-to-next generation 
of $\beta\beta0\nu$ decay experiments, $10^{-2}$ eV. This is so, independently of the light neutrino hierarchy.

Finally, there is a still plausible possibility of having a significant contribution to the $\beta\beta0\nu$ decay
from a sub-eV  thermal sterile neutrino which can satisfy the cosmological bounds. For example, if $f_{s_1}(T_{\rm max})\geq 1$
with $M_1\lesssim1$eV and $M_{2,3}\gg100$MeV, the lightest sterile neutrino could give a significant contribution
to the process. However, for such a low $M_1$ scale,  the constraints from neutrino oscillations are expected to be 
very relevant. Therefore, this case deserves a more careful analysis which should also face the possibility of explaining the 
neutrino anomalies. This would also apply to  the scenario where $M_1\leq 1$eV, 1 eV $\leq M_2\leq 100$ MeV and $M_3 \gg$ 
100 MeV, if $m_1 \leq m_1^{th}$. The two lighter states would contribute to $\beta\beta0\nu$. The contribution of $M_2$ would 
be similar to that of $M_1$ in Fig.~\ref{fig:mbb}, while that of $M_1$ would depend significantly on oscillation constraints. 

\section{Conclusions}
\label{sec:conclusions}

We have studied the thermalization of the heavy sterile neutrinos in the standard Type I seesaw model 
with three extra singlets and a low-scale, $eV \leq M_j \leq 100$MeV. 
The production of the states in the Early Universe occurs via non-resonant mixing (in the absence of 
large primordial asymmetries) and we have found that, independently of the unknown mixing parameters 
in the model, full thermalization is always reached for the three states if the lightest neutrino mass is above ${\mathcal O}(10^{-3}$eV). Since, they decouple 
early, while they are still relativistic, these states either violate BBN constraints on $\Delta N_{\rm eff}$ or/and 
contribute too much energy density to the Universe at later times, either in the form of cold dark matter (if they decay 
late enough) or in the form of dark radiation (if they decay earlier). Majorana masses would all need to be heavier than 
${\mathcal O}(100$MeV)  to avoid cosmology constraints, or alternatively one of them could remain very light sub-eV, 
resulting in a milder tension with cosmology. 

In contrast, if the lightest neutrino mass is below ${\mathcal O}(10^{-3}$eV), one and only one of the sterile states might 
never thermalize, depending on the unknown  parameters of the model, and therefore its mass is unconstrained. The other two 
states always thermalize and therefore their masses should be above ${\mathcal O}(100$MeV) to avoid cosmological constraints. 
The scenario often referred to as the $\nu$MSM \cite{Asaka:2005pn} falls in this category, where the non-thermalized state
in the keV region could be a candidate for warm dark matter \cite{Dodelson:1993je,Asaka:2005an} and the heavier states 
could generate the baryon asymmetry \cite{Akhmedov:1998qx}. In fact, a more stringent upper bound on $m_1$ had been previously 
derived from the requirement that $M_1 \sim$ keV and could be a warm dark matter candidate \cite{Asaka:2005pn}. Alternatively, 
the tension with cosmology could also be minimized in this case if one of the two thermalized states is very light sub-eV and 
the other remains heavy. 

Although the possibility of having one of the species in the sub-eV range could provide an 
interesting scenario to maybe explain the neutrino oscillation anomalies, the tension between 
cosmology and neutrino oscillation experiments is likely to be significant. 

Finally, we have also studied the impact of the cosmological bounds extracted in this work on the $\beta\beta0\nu$ decay 
phenomenology. We have found that if one of the sterile neutrinos does not thermalize, the quasi-degenerate light neutrino 
spectrum would be ruled out. The region of the parameter space in which a cancellation can take place in the active neutrino 
contribution is also excluded in this scenario. In addition, we have also shown that the contributions of sterile states 
with $M_1 \in [1$eV-$100$MeV$]$ are subleading and well beyond the sensitivity of the next-to-next generation of 
$\beta\beta0\nu$ decay experiments. However, a sub-eV thermal sterile state could give a contribution, in this scenario, 
within reach of the next-to-next generation of $\beta\beta0\nu$ decay experiments, the constraints from neutrino 
oscillations playing a very important role.

\begin{acknowledgments}

We thank M. Laine for providing us with data files with the numerical results of ref.~\cite{Asaka:2006nq} in an easy-to-use format. We also thank
A.~Donini,  E.~Fern\'andez-Mart\'{\i}nez, O.~Mena, S.~Petcov and A.Vincent for useful discussions.  
This work was partially supported by grants FPA2011-29678, PROMETEO/2009/116, CUP (CSD2008-00037), 
ITN INVISIBLES (Marie Curie Actions, PITN-GA-2011-289442) and the INFN program on 
``Astroparticle Physics.'' J. L\'opez-Pav\'on acknowledges NORDITA for the hospitality during the ``News in Neutrino Physics'' workshop, where 
part of this work was done. M. Kekic thanks the IFT-UAM/CSIC institute for hospitality while this work was done.

\end{acknowledgments}

\appendix 

\section{Appendix }
\label{sec:appendix}

In the density matrix formalism  \cite{Dolgov:2002wy}, the kinetic equations have the usual form:
\begin {align}
 \dot{\rho}=-i [H,\rho]-\frac{1}{2}\{\Gamma,\rho-\rho_{eq}I_A\};\label{Boltzman}
\end {align}
where $\rho$ is $6 \times 6$ density matrix, $H$ is the Hamiltonian describing the propagation of relativistic neutrinos 
in the plasma, $\Gamma$ is the collision term that we take from refs.~\cite{Asaka:2006nq,laine}, 
and  $\rho_{eq}$ is the active neutrino thermal density, i.e. the Fermi-Dirac distribution $\rho_{eq}=\frac{1}{e^{E/T}+1}$, in the absence of a chemical potential. $I_A$ is the projector 
on the active sector.  The trace of the density matrix corresponds to the number density of neutrinos.

Rewriting eq.~(\ref{Boltzman}) in the form of active-sterile block matrices we get the following set of equations:
\begin{eqnarray}
 \dot{\rho}_A &=& -i (H_A\rho_A-\rho_AH_A+H_{AS}\rho^\dagger_{AS}-\rho_{AS}H^{\dagger}_{AS})-\frac{1}{2}\{\Gamma_A, \rho_A-\rho_{eq}I_A\},\label{rhoaa}\\
 \dot{\rho}_{AS}&=&-i(H_A\rho_{AS}-\rho_A H_{AS}+H_{AS}\rho_S-\rho_{AS}H_S)-\frac{1}{2}\Gamma_A\rho_{AS}, \label{rhoas} \\
 \dot{\rho}_S &=& -i (H^\dagger_{AS}\rho_{AS}-\rho^\dagger_{AS}H_{AS}+H_S\rho_S-\rho_S H_S). \label{rhos}
 \end{eqnarray}

Assuming that $\Gamma_A \gg$ Hubble rate, we can approximate 
\begin{eqnarray}
\dot\rho_A = \dot\rho_{AS} = 0.
\end{eqnarray}
This is the so-called ``static approximation'' \cite{Foot:1996qc, Bell:1998ds,Dolgov:2003sg}. 

The first equation implies  $\rho_A=\rho_{eq} I_A$, while the second equality implies 
\begin{eqnarray}
(\rho_{AS})_{ai} = ( - (H_A - \tilde{H}_i I_A) + i \Gamma_A /2)^{-1}_{aa'}  (H_{AS})_{a'j} ((\rho_S)_{ji}- \rho_{eq} \delta_{ji}), 
\end{eqnarray}
where we have made the approximation that $(H_S)_{ij} = \tilde{H}_i \delta_{ij}$, which is very good in the seesaw limit. 
Similarly we find
\begin{eqnarray}
(\rho^\dagger_{AS})_{ia} = ((\rho_S)_{ij}- \rho_{eq} \delta_{ij})  (H^\dagger_{AS})_{ja'}  ( - (H_A - \tilde{H}_i I_A) - i \Gamma_A/2 )^{-1}_{a'a}  
\end{eqnarray}
Defining $\tilde{\rho}_S \equiv \rho_S - \rho_{eq} I_S$, and after substituting 
$\rho_{AS}$ and $\rho^\dagger_{AS}$ in eq. (\ref{rhos}), we get the following equation

\begin{eqnarray}
(\dot\rho_S)_{ij} &=&  - i (\tilde{H}_iÊ- \tilde{H}_j) (\rho_S)_{ij} \nonumber\\
& &- i (H_{AS}^\dagger)_{a'i} ( - (H_A - \tilde{H}_j I_A) + i \Gamma_A/2 )^{-1}_{a'a} (H_{AS})_{ak} \tilde{\rho}_{kj}\nonumber\\
&& +i \tilde{\rho}_{ik} (H_{AS}^\dagger)_{a'k} ( - (H_A - \tilde{H}_i I_A) - i \Gamma_A/2 )^{-1}_{a'a} (H_{AS})_{aj}.
\label{eq:rhosij}
\end{eqnarray}
It is clear that the equilibrium distribution for the sterile components is $\tilde{\rho}_{ii} = 0$ or $\rho_{ii} = \rho_{eq} \delta_{ii}$. 

At this point it is necessary to solve the $3\times 3$ system of differential equations eqs.~(\ref{eq:rhosij}), but we can further simplify the problem if we assume that the dynamics of the different sterile components decouple from each other, which is the case provided their masses are sufficiently different. Since $H_{AS}$ depends
on temperature, if the sterile splittings are significantly different from each other, we will generically have that $H_{AS}$ will be very suppressed
unless the temperature-dependent effective mass is similar to one of the mass splittings. Let us suppose that this is the case. At high $T$ all 
active-sterile mixings are very suppressed, until one splitting that associated to the sterile state $s$ is reached; at this point only $(H_{AS})_{as}$ is non-negligible. 
Then only $(\rho_S)_{ss}$ changes significantly and can be described by
\begin{eqnarray}
\dot{\rho}_{ss} &=& - i \left(H_{AS}^\dagger \left\{\frac{1}{- (H_A- \tilde{H}_s) +  i \Gamma_A/2}- \frac{1}{- (H_A- \tilde{H}_s) - i \Gamma_A/2} \right\}H_{AS}\right)_{ss} \tilde\rho_{ss} \nonumber\\
&=& -   \left(H_{AS}^\dagger \left\{\frac{\Gamma_A}{(H_A- \tilde{H}_s)^2 +   \Gamma^2_A/4}  \right\}H_{AS}\right)_{ss} \tilde\rho_{ss},
\label{eq:rhoss}
\end{eqnarray}
where in the last step we have assumed that $H_A, \Gamma_A$ commute, which again is a good approximation in the seesaw limit. This equation justifies eq.~(\ref{eq:gammas}), since the source term on the right of eq.~(\ref{eq:rhoss}) is the same as $\Gamma_{s}$ in eq.~(\ref{eq:gammas}) if we neglect the term $\sim \Gamma_A^2$ in the denominator.   
We have checked that the result of solving the three coupled equations or the three independent ones give very similar results and the latter is obviously much faster. 

Now we have to consider the evolution in an expanding Universe, where the variation of the scale factor $a(t)$, depends on the Hubble expansion rate, which, in a 
radiation-dominated Universe at temperature $T$, is given by 
\begin{eqnarray}
H(T) = \sqrt{ \frac{8 \pi G_N}{3} \left( \frac{\pi^2}{30} g_*(T) T^4 +  \epsilon_s(T)\right)},
\end{eqnarray}
where $g_*$ counts the relativistic degrees of freedom and we have included the contribution to the energy density of the sterile states,  $\epsilon_s$, which must be computed integrating 
 the trace of the density matrix, $\rho_S$.  
As in ref.~\cite{Dolgov:2003sg} we introduce new variables:
\begin{align}
 x= \frac{a(t)}{a_{BBN}} , \;\;\;  y=x \frac{p}{T_{BBN}} ;
\end{align}
where  $a(t)$ is cosmic scale factor, $T_{BBN}\simeq 1$MeV is the temperature of active neutrino decoupling and 
$a_{BBN}$ the scale factor at this point. On other hand, entropy conservation implies $g_{S*}(T) T^3 a(t)^3$=constant (here $g_{S*}$ refers to the relativistic degrees of freedom in equilibrium, it differs from $g_*$ in the Hubble expansion only after light neutrino decoupling). 
This relation implies
\bea
x = \frac{T_{BBN}}{T} \left(\frac{g_{S*}(T_{BBN})}{g_{S*}(T)}\right)^{1/3}.
\eea
 We neglect the contribution of the sterile states to $g_{S*}$, because they decouple very early and therefore they give a small contribution. 

The time derivative acting on any phase space distribution can be written as:
\begin{align}
 \frac{d}{dt}f(t,p)=(\partial_t-Hp\partial_p)f(t,p)=Hx\partial_xf(x,y).
\end{align}
Applied to eq.~(\ref{Boltzman}) this leads to 
\begin{align}
\left. H x \frac{\partial}{\partial x} \rho(x,y)\right|_{y}=-i[\hat{H},\rho(x,y)]-\frac{1}{2}\{\Gamma,\rho(x,y)-\rho_{eq}(x,y)I_A\},
\end{align}
where 
\begin{eqnarray}
\rho_{eq}(x,y) = \frac{1}{\exp\left[ y (g_{S*}(T(x))/g_{S*}(T_{BBN}) )^{1/3}\right] +1},
\end{eqnarray}
and for eq.~(\ref{eq:rhoss}) similarly
\begin{align}
\left. H x \frac{\partial}{\partial x} \rho_{ss}(x,y)\right|_{y}= -   \left(H_{AS}^\dagger \left\{\frac{\Gamma_A}{(H_A- \tilde{H}_s)^2 +   \Gamma^2_A/4}  \right\}H_{AS}\right)_{ss} \tilde\rho_{ss}(x,y),
\end{align}

The equations are evolved  from an initial condition  at $x_i \rightarrow 0$, $\rho_{ss}=0$,  until active neutrino decoupling, $x_f=1$ for fixed $y$. 
We define the effective number of additional neutrino species by
\begin{align}
 \Delta N_{\rm eff}=\frac{\epsilon_s}{\epsilon_\nu^0},
\end{align}
where  $\epsilon_\nu^0$ is the energy density of one SM massless neutrino. For each additional  neutrino we compute the 
contribution to $\Delta N_{\rm eff}$ from the solution of $\rho_{s_js_j}(x_f,y)$ as
\begin{equation}
\Delta N^{(j) BBN}_{\rm eff}|_{energy}=\frac{\int \mathrm{d}y ~y^2 E(y) \rho_{s_j s_j}(x_f,y)}{\int \mathrm{d}y ~y^2 p(y) \rho_{eq}(x_f,y)},
\label{eq:dneffen}
\end{equation}
where $p(y)=\frac{y}{x_f} T_{BBN}$ and $E(y)=\sqrt{p(y)^2+M_j^2}$.

We can also define the ratio of number densities instead, which is more appropriate when they are not relativistic,
\begin{equation}
 \Delta N^{(j) BBN}_{\rm eff}|_{number}=\frac{\int \mathrm{d}y ~y^2 \rho_{s_j s_j}(x_f,y)}{\int \mathrm{d}y~ y^2  \rho_{eq}(x_f,y)},
 \label{eq:dneffnum}
\end{equation} 
The two correspond to the solid/dashed curves depicted in Fig.~\ref{fig:non thermal}.


\vspace{0.5cm}
\begin{thebibliography}{99}

    
 \bibitem{seesaw}
 P. Minkowski, Phys.\ Lett. \ B {\bf67}, 421 (1977); M. Gell-Mann, P. Ramond and R. Slansky, in {\it Supergravity}, edited by P. van
 Nieuwenhuizen and D. Freedman (North-Holland, 1979), p. 315; T. Yanagida in {\it Proceedings of the Workshop on the Unified Theory and 
 the Baryon Number in the Universe}, edited by O. Sawada and A. Sugamoto (KEK Report No. 79-18, Tsukuba, 1979), p. 95; R.N. 
 Mohapatra and G. Senjanovi
c, Phys. Rev. Lett. {\bf 44},  912 (1980).
    
      \bibitem{miniseesaw}
  A.~de Gouvea,
  Phys.\ Rev.\ D {\bf 72}, 033005 (2005).
  A.~de Gouvea, J.~Jenkins and N.~Vasudevan,
  Phys.\ Rev.\ D {\bf 75}, 013003 (2007).
  
    \bibitem{Donini:2011jh}
  A.~Donini {\it et al},
  JHEP {\bf 1107}, 105 (2011).


    \bibitem{Aguilar:2001ty}
  A.~Aguilar-Arevalo {\it et al.}  [LSND Collaboration],
  Phys.\ Rev.\ D {\bf 64}, 112007 (2001).
    
    \bibitem{Aguilar-Arevalo:2013pmq}
  A.~A.~Aguilar-Arevalo {\it et al.}  [MiniBooNE Collaboration],
  Phys.\ Rev.\ Lett.\  {\bf 110}, 161801 (2013).
  
  \bibitem{reactor}
  G.~Mention {\it et al}, 
  Phys.\ Rev.\ D {\bf 83}, 073006 (2011).
  P.~Huber,
  Phys.\ Rev.\ C {\bf 84}, 024617 (2011)
   [Erratum-ibid.\ C {\bf 85}, 029901 (2012)].
  
   \bibitem{Dodelson:1993je}
  S.~Dodelson and L.~M.~Widrow,
  Phys.\ Rev.\ Lett.\  {\bf 72}, 17 (1994).

\bibitem{Shi:1998km}
  X.~-D.~Shi and G.~M.~Fuller,
  Phys.\ Rev.\ Lett.\  {\bf 82}, 2832 (1999)
  [astro-ph/9810076].

\bibitem{Abazajian:2001nj}
  K.~Abazajian, G.~M.~Fuller and M.~Patel,
  Phys.\ Rev.\ D {\bf 64}, 023501 (2001)
  [astro-ph/0101524].
  
\bibitem{Asaka:2005an}
  T.~Asaka, S.~Blanchet and M.~Shaposhnikov,
  Phys.\ Lett.\ B {\bf 631}, 151 (2005)
  [hep-ph/0503065].
  

\bibitem{Bulbul:2014sua}
  E.~Bulbul {\it et al.}, 
  arXiv:1402.2301 [astro-ph.CO].
  
\bibitem{Boyarsky:2014jta} 
  A.~Boyarsky, O.~Ruchayskiy, D.~Iakubovskyi and J.~Franse,
  arXiv:1402.4119 [astro-ph.CO].
  
\bibitem{Akhmedov:1998qx}
  E.~K.~Akhmedov, V.~A.~Rubakov and A.~Y.~Smirnov,
  Phys.\ Rev.\ Lett.\  {\bf 81}, 1359 (1998).

\bibitem{Asaka:2005pn}
  T.~Asaka and M.~Shaposhnikov,
  Phys.\ Lett.\ B {\bf 620}, 17 (2005)
  [hep-ph/0505013].
  
  

  
\bibitem{Canetti:2012kh}
  L.~Canetti {\it et al},
  Phys.\ Rev.\ D {\bf 87}, 093006 (2013) and references therein.
    
     \bibitem{Abada:2007ux}
 S.~Antusch, C.~Biggio, E.~Fernandez-Martinez, M.~B.~Gavela and J.~Lopez-Pavon,
 JHEP {\bf 0610}, 084 (2006)
 [hep-ph/0607020].
 A.~Abada, C.~Biggio, F.~Bonnet, M.~B.~Gavela and T.~Hambye,
 JHEP {\bf 0712}, 061 (2007)
 [arXiv:0707.4058 [hep-ph]].  
 S.~Antusch, J.~P.~Baumann and E.~Fernandez-Martinez,
 Nucl.\ Phys.\ B {\bf 810}, 369 (2009)
 [arXiv:0807.1003 [hep-ph]].
  D.~N.~Dinh, A.~Ibarra, E.~Molinaro and S.~T.~Petcov,
 JHEP {\bf 1208}, 125 (2012)
  [Erratum-ibid.\  {\bf 1309}, 023 (2013)]
 [arXiv:1205.4671 [hep-ph]].
R.~Alonso, M.~Dhen, M.~B.~Gavela and T.~Hambye,
 JHEP {\bf 1301}, 118 (2013)
 [arXiv:1209.2679 [hep-ph]].
 D.~N.~Dinh and S.~T.~Petcov,
 JHEP {\bf 1309}, 086 (2013)
 [arXiv:1308.4311 [hep-ph]].

\bibitem{shaposhnikov}
D. Gorbunov and M. Shaposhnikov, JHEP {\bf 10}, 015 (2007).

\bibitem{Atre:2009rg}
  A.~Atre, T.~Han, S.~Pascoli and B.~Zhang,
  JHEP {\bf 0905}, 030 (2009).

\bibitem{Foot:1995qk} 
  R.~Foot, M.~J.~Thomson and R.~R.~Volkas,
  Phys.\ Rev.\ D {\bf 53}, 5349 (1996)
  [hep-ph/9509327].
  
\bibitem{Hannestad:2013ana} 
  S.~Hannestad, R.~S.~Hansen and T.~Tram,
  Phys.\ Rev.\ Lett.\  {\bf 112}, no. 3, 031802 (2014)
  [arXiv:1310.5926 [astro-ph.CO]].
  
\bibitem{Dasgupta:2013zpn} 
  B.~Dasgupta and J.~Kopp,
  Phys.\ Rev.\ Lett.\  {\bf 112}, no. 3, 031803 (2014)
  [arXiv:1310.6337 [hep-ph]].

\bibitem{Cooke:2013cba}
  R.~Cooke, M.~Pettini, R.~A.~Jorgenson, M.~T.~Murphy and C.~C.~Steidel,
  arXiv:1308.3240 [astro-ph.CO].


  \bibitem{Ade:2013zuv}
  P.~A.~R.~Ade {\it et al.}  [Planck Collaboration],
  arXiv:1303.5076 [astro-ph.CO].
 
 \bibitem{WMAP}
  G.~Hinshaw {\it et al.}  [WMAP Collaboration],
  Astrophys.\ J.\ Suppl.\  {\bf 208}, 19 (2013).
  
  
\bibitem{SPT}
  Z.~Hou {\it et al.},
  arXiv:1212.6267 [astro-ph.CO].


\bibitem{ACT}  
  J.~L.~Sievers {\it et al}, 
  arXiv:1301.0824 [astro-ph.CO].
  
\bibitem{Ade:2014xna} 
  P.~A.~R.~Ade {\it et al.}  [BICEP2 Collaboration],
  Phys.\ Rev.\ Lett.\  {\bf 112}, 241101 (2014)
  [arXiv:1403.3985 [astro-ph.CO]].
  
\bibitem{Ade:2014gua} 
  P.~A.~R.~Ade {\it et al.}  [BICEP2 Collaboration],
  Astrophys.\ J.\  {\bf 792}, 62 (2014)
  [arXiv:1403.4302 [astro-ph.CO]].
\bibitem{Giusarma:2014zza}
  E.~Giusarma, E.~Di Valentino, M.~Lattanzi, A.~Melchiorri and O.~Mena
  arXiv:1403.4852 [astro-ph.CO].
  
\bibitem{Archidiacono:2014apa} 
  M.~Archidiacono {\it et al.},
  JCAP {\bf 1406}, 031 (2014)
  [arXiv:1404.1794 [astro-ph.CO]].
  
\bibitem{Bergstrom:2014fqa} 
  J.~Bergström, M.~C.~Gonzalez-Garcia, V.~Niro and J.~Salvado,
  arXiv:1407.3806 [hep-ph].

  \bibitem{Dolgov:2003sg}
  A.~D.~Dolgov and F.~L.~Villante,
    Nucl.\ Phys.\ B {\bf 679}, 261 (2004).
  
  \bibitem{Cirelli:2004cz}
  M.~Cirelli, G.~Marandella, A.~Strumia and F.~Vissani,
    Nucl.\ Phys.\ B {\bf 708}, 215 (2005).
  
    \bibitem{Melchiorri:2008gq}
  A.~Melchiorri {\it et al},
    JCAP {\bf 0901}, 036 (2009).

 \bibitem{Hannestad:2012ky}
  S.~Hannestad, I.~Tamborra and T.~Tram,
    JCAP {\bf 1207}, 025 (2012).

  \bibitem{Kuflik:2012sw}
  E.~Kuflik, S.~D.~McDermott and K.~M.~Zurek,
    Phys.\ Rev.\ D {\bf 86}, 033015 (2012).

\bibitem{Mirizzi:2012we} 
  A.~Mirizzi, N.~Saviano, G.~Miele and P.~D.~Serpico,
  Phys.\ Rev.\ D {\bf 86}, 053009 (2012)
  [arXiv:1206.1046 [hep-ph]].
  
  \bibitem{Jacques:2013xr}
  T.~D.~Jacques, L.~M.~Krauss and C.~Lunardini,
    Phys.\ Rev.\ D {\bf 87}, 083515 (2013).
 
\bibitem{Saviano:2013ktj} 
  N.~Saviano {\it et al.}, 
  Phys.\ Rev.\ D {\bf 87}, 073006 (2013)
  [arXiv:1302.1200 [astro-ph.CO]].
  
\bibitem{Archidiacono:2013xxa} 
  M.~Archidiacono, N.~Fornengo, C.~Giunti, S.~Hannestad and A.~Melchiorri,
  Phys.\ Rev.\ D {\bf 87}, no. 12, 125034 (2013)
  [arXiv:1302.6720 [astro-ph.CO]].
   
\bibitem{Mirizzi:2013kva} 
  A.~Mirizzi {\it et al.},
  Phys.\ Lett.\ B {\bf 726}, 8 (2013)
  [arXiv:1303.5368 [astro-ph.CO]].

\bibitem{Hernandez:2013lza} 
  P.~Hernandez, M.~Kekic and J.~Lopez-Pavon,
  Phys.\ Rev.\ D {\bf 89}, 073009 (2014)
  [arXiv:1311.2614 [hep-ph]].

\bibitem{Shaposhnikov:2008pf} 
  M.~Shaposhnikov,
  JHEP {\bf 0808}, 008 (2008)
  [arXiv:0804.4542 [hep-ph]].
  
  
  \bibitem{Casas:2001sr}
  J.~A.~Casas and A.~Ibarra,
  Nucl.\ Phys.\ B {\bf 618}, 171 (2001).

  \bibitem{Donini:2012tt}
  A.~Donini {\it et al},
  JHEP {\bf 1207}, 161 (2012).

 \bibitem{Blennow:2011vn}
  M.~Blennow and E.~Fernandez-Martinez,
  Phys.\ Lett.\ B {\bf 704}, 223 (2011).

   \bibitem{simple:1990} R.~Barbieri and A.~Dolgov,
  Phys.\ Lett.\ B {\bf 237}, 440  (1990) {and Nucl. Phys. B349 (1991) 743}.  K.~Kainulainen,
  Phys.\ Lett.\ B {\bf 244}, 191 (1990). 
 
 \bibitem{Notzold:1987ik}
  D.~Notzold and G.~Raffelt,
  Nucl.\ Phys.\ B {\bf 307}, 924 (1988). 
  
  
   \bibitem{oldies}
  L. Stodolsky, Phys.\ Rev.\ D{\bf 36} (1987) 2273. G. Raffelt, G. Sigl and L. Stodolsky, Phys. \ Rev.\ Lett.\ {\bf 70},
  2363 (1993). M.J. Thomson, Phys. Rev. A {\bf 45}, 2243 (1992). 

 \bibitem{polariz}  K. Enqvist, K. Kainulainen, M. Thomson, Nucl.\ Phys.\ B373 498 (1992). B.H.J McKellar and M.J. Thomson, 
 Phys.\ Rev.\ D {\bf 49}, 2710 (1994). 
 
   \bibitem{Sigl:1992fn}
  G.~Sigl and G.~Raffelt,
  Nucl.\ Phys.\ B {\bf 406}, 423 (1993). 
  
  
\bibitem{Dolgov:2002wy}
  A.~D.~Dolgov,
  Phys.\ Rept.\  {\bf 370}, 333  (2002)
  [hep-ph/0202122].
   
   
     \bibitem{Abazajian:2002yz}
  K.~N.~Abazajian and G.~M.~Fuller,
    Phys.\ Rev.\ D {\bf 66}, 023526 (2002). 
  K.~Abazajian,
  Phys.\ Rev.\ D {\bf 73}, 063506 (2006).  
  
    \bibitem{Asaka:2006nq}
  T.~Asaka, M.~Laine and M.~Shaposhnikov,
  JHEP {\bf 0701}, 091 (2007).
  
  \bibitem{laine} Data files for the imaginary part of the neutrino self-energy have been made available in http://www.laine.itp.unibe.ch/neutrino-rate/.
  
\bibitem{KolbTurner} 
E.~W.~Kolb and M.~S.~Turner,
The Early Universe,
Front.\ Phys.\  {\bf 69}, 1 (1990).

\bibitem{Dolgov:2000pj}
  A.~D.~Dolgov, S.~H.~Hansen, G.~Raffelt and D.~V.~Semikoz,
  Nucl.\ Phys.\ B {\bf 580}, 331 (2000) and   Nucl.\ Phys.\ B {\bf 590}, 562 (2000).

  \bibitem{Ruchayskiy:2012si}
  O.~Ruchayskiy and A.~Ivashko,
  JCAP {\bf 1210}, 014 (2012).


\bibitem{Ruchayskiy:2011aa}
  O.~Ruchayskiy and A.~Ivashko,
  JHEP {\bf 1206}, 100 (2012).

\bibitem{GonzalezGarcia:2012yq} 
  M.~C.~Gonzalez-Garcia, V.~Niro and J.~Salvado,
  JHEP {\bf 1304}, 052 (2013)
  [arXiv:1212.1472 [hep-ph]].

\bibitem{Hasenkamp:2012ii} 
  J.~Hasenkamp and J.~Kersten,
  JCAP {\bf 1308}, 024 (2013)
  [arXiv:1212.4160 [hep-ph]].

\bibitem{Hasenkamp:2014hma} 
  J.~Hasenkamp,
  arXiv:1405.6736 [astro-ph.CO].
  

\bibitem{Laine:2008pg}
  M.~Laine and M.~Shaposhnikov,
  JCAP {\bf 0806}, 031 (2008)
  [arXiv:0804.4543 [hep-ph]].
      
      
\bibitem{Foot:1996qc}
  R.~Foot and R.~R.~Volkas,
  Phys.\ Rev.\ D {\bf 55}, 5147 (1997).
  
  \bibitem{Bell:1998ds}
  N.~F.~Bell, R.~R.~Volkas and Y.~Y.~Y.~Wong,
  Phys.\ Rev.\ D {\bf 59}, 113001 (1999).
  
\bibitem{Boyarsky:2008ju} 
  A.~Boyarsky, O.~Ruchayskiy and D.~Iakubovskyi,
  JCAP {\bf 0903}, 005 (2009)
  [arXiv:0808.3902 [hep-ph]].
  
\bibitem{Blennow:2010th}
  M.~Blennow, E.~Fernandez-Martinez, J.~Lopez-Pavon and J.~Menendez,
  JHEP {\bf 1007}, 096 (2010)
  [arXiv:1005.3240 [hep-ph]].

\bibitem{LopezPavon:2012zg}
  J.~Lopez-Pavon, S.~Pascoli and C.~-f.~Wong,
  Phys.\ Rev.\ D {\bf 87}, 9 (2013),  093007
  [arXiv:1209.5342 [hep-ph]].
  





\end{thebibliography}

\providecommand{\href}[2]{#2}\begingroup\raggedright

\endgroup

\end{document}